\documentclass[onecolumn,nonote,noversion]{cdcarticle}
\def\MODE{1}

\usepackage{math}
\usepackage{tikz}
\usepackage{enumerate}
\usepackage[hidelinks]{hyperref}

\newcommand{\cl}{\mathcal{T}_{c\ell}}
\newcommand{\PROB}{\textup{\hyperref[PROB]{\bf METP}}}
\newcommand{\ITS}{\textup{\hyperref[ITERATE]{\bf ITS}}}
\newcommand{\shrink}[1]{\addtolength{\arraycolsep}{-#1 pt}}
\DeclareMathOperator{\ord}{\mathcal{O}}
\DeclareMathOperator{\ess}{\textrm{ess}}
\DeclareMathOperator*{\esssup}{\ess\sup}
\newcommand{\bref}[1]{\hyperref[#1]{\bf(#1)}}	

\usetikzlibrary{matrix,fit,backgrounds,plotmarks,calc}
\usepackage{pgfplots}

\if\MODE1
	\pgfplotsset{width=12cm,height=9cm,compat=newest}
\else
	\pgfplotsset{width=9cm,height=7cm,compat=newest}
\fi

\tikzset{dashdot/.style={dash pattern=on 1pt off 2.5pt on 4.5pt off 2.5pt}}
\pgfplotscreateplotcyclelist{gray marks}{%
every mark/.append style={fill=gray,scale=0.8},mark=*\\%
every mark/.append style={fill=gray},mark=diamond*\\%
every mark/.append style={fill=gray},mark=triangle*\\%
every mark/.append style={fill=gray,scale=0.8},mark=square*\\%
}
\pgfplotscreateplotcyclelist{color dashes}{%
blue,thick\\
blue,thick,dashed\\
red,very thick,dashdot\\
purple!40!white,thick\\
purple!40!white,thick,dashed\\
}
\pgfplotscreateplotcyclelist{gray lines}{%
black,very thin\\%
}

\pgfplotsset{
  tick label style = {font=\footnotesize},
  every axis label = {font=\footnotesize},
  legend style = {font=\footnotesize},
  label style = {font=\small}
}

\begin{document}

\title{State-space solution to a minimum-entropy $\mathcal{H}_\infty$-optimal control problem with a nested information constraint}

\if\MODE1\author{Laurent Lessard}
\else\author{Laurent Lessard\footnotemark[1]}\fi

\note{To Appear, IEEE Conference on Decision and Control}
\maketitle

\if\MODE1\else
\footnotetext[1]{L.~Lessard is with the Department of Mechanical Engineering at the University of California, Berkeley, CA~94720, USA.\\ \texttt{lessard@berkeley.edu}}
\fi


\begin{abstract}
State-space formulas are derived for the minimum-entropy $\Hinf$ controller when the plant and controller are constrained to be block-lower-triangular. Such a controller exists if and only if: the corresponding unstructured problem has a solution, a certain pair of coupled algebraic Riccati equations admits a mutually stabilizing fixed point, and a pair of spectral radius conditions is met. The controller's observer-based structure is also discussed, and a simple numerical approach for solving the coupled Riccati equations is presented.
\end{abstract}


\section{Introduction}\label{sec:intro}

Entropy minimization may be thought of as link between the popular $\Htwo$ and $\Hinf$ performance measures. Given a stable matrix transfer function (MTF) $\mathcal{F}(s)$, the entropy with tolerance $\gamma > 0$ is defined as
\begin{equation}\label{def:entropy}
\mathcal{I}_\gamma( \mathcal{F}) \defeq \\
-\frac{\gamma^2}{2\pi} \int_{-\infty}^\infty \log \left| \det\bigl(I - \gamma^{-2}  \mathcal{F}(j\omega)^* \mathcal{F}(j\omega)\bigr) \right| d\omega
\end{equation}
A key property of entropy is that $\mathcal{I}_\gamma( \mathcal{F} )$ is finite if and only if $\|\mathcal{F}\|_\infty < \gamma$. So entropy may be used as a surrogate for the $\Hinf$ norm when searching for suboptimal controllers with a prescribed performance $\gamma$. Entropy is also related to the $\Htwo$ performance measure in the limit  $\lim_{\gamma\to\infty}\mathcal{I}_\gamma( \mathcal{F} ) = \|\mathcal{F}\|_2^2$.

The focus of this paper is a structurally constrained version of the standard $\Hinf$ control problem whereby a given sparsity pattern is imposed on the controller. Such constraints arise in decentralized control; rows of the controller MTF $\mathcal{K}$ may be thought of as separate controllers and the constraint $\mathcal{K}_{ij}=0$ means that controller $i$ does not measure measurement $j$.

The plant and controller are continuous linear time-invariant systems described by the equations
\[
\bmat{z \\ y} = \bmat{\mathcal P_{11} & \mathcal P_{12} \\ \mathcal P_{21} &  \mathcal P_{22} } \bmat{w \\ u}
\qquad\text{and}\qquad
u =  \mathcal K y
\]
where $z$ and $w$ are the regulated output and exogenous input, and $y$ and $u$ are the measurement and controlled input, respectively. $\mathcal{P}_{22}$ and $\mathcal{K}$ are assumed to each have a $2\times 2$ block-lower-triangular sparsity structure with conforming dimensions. That is:
\begin{equation}\label{triangular_structure}
\mathcal{P}_{22} \defeq \bmat{ \mathcal G_{11} & 0 \\ \mathcal G_{21} & \mathcal G_{22} }
\quad\text{and}\quad
\mathcal K \defeq \bmat{ \mathcal K_{11} & 0 \\ \mathcal K_{21} & \mathcal K_{22} }
\end{equation}
where $\mathcal{G}_{ij}$ is a strictly proper rational $k_i\times m_j$ MTF and $\mathcal{K}_{ij}$ is a proper rational $m_i\times k_j$ MTF. The closed-loop map $w\to z$ found by eliminating $\mathcal{K}$ is given by
\begin{equation}\label{closed-loop_map}
\cl \defeq \mathcal P_{11} + \mathcal P_{12} \mathcal K ( I - \mathcal P_{22} \mathcal K )^{-1} \mathcal P_{21}
\end{equation}
The main problem statement is given below.

\paragraph{Min-entropy $\Hinf$ two-player problem (\PROB)}\label{PROB}\hfill\\
Given a plant $\mathcal{P}$ defined as above, find a controller $\mathcal{K}$ satisfying the following four requirements
\begin{enumerate}[\bf({R}1)]
\itemsep=1mm
\item $\mathcal{K}$ stabilizes $\mathcal{P}$. \label{R1}
\item The closed-loop map satisfies $\|\cl\|_\infty < \gamma $. \label{R2}
\item The entropy $\mathcal{I}_\gamma(\cl)$ is minimized. \label{R3}
\item $\mathcal{K}$ has the triangular structure~\eqref{triangular_structure}. \label{R4}
\end{enumerate}

Requirements \bref{R1}--\bref{R2} describe the standard $\Hinf$ control problem. Existing approaches include the seminal work by Doyle, Glover, Kargonekar, and Francis (DGKF)~\cite{dgkf}, and the linear matrix inequality (LMI) approach by Gahinet and Apkarian~\cite{gahinet_apkarian}. It was shown that the DGKF controller also minimizes entropy~\cite{glover_mustafa_entropy}, as in \bref{R3}. The risk-sensitive approach of Whittle~\cite{whittle} is also closely related to the entropy formulation. In the limit $\gamma \to \infty$, \bref{R1}--\bref{R3} reduce to the standard $\Htwo$ optimal control problem. These various approaches and their interpretations are well covered in many modern texts on robust control. See for example \cite{dullerud,zdg}.

The structural constraint \bref{R4} complicates the problem substantially. While the associated optimization may still be convexified~\cite{voulgaris_stabilization,rotkowitz06}, it remains infinite-dimensional and there is no obvious way to find closed-form solutions.
Nevertheless, \PROB~was solved in the limiting $\Htwo$ case by Lessard and Lall~\cite{lessard_lall_tpof,lessard_tac}. In related work~\cite{shah_parrilo,swigart10}, the $\Htwo$ case was solved under the further assumption of noise-free state measurements (full-state feedback). 

Unlike the limiting $\Htwo$ case, \PROB~for a general~$\gamma$ does not become appreciably simpler under full-state feedback assumptions. The only existing solution to the general~\PROB~is by Scherer~\cite{scherer}, and uses an LMI approach reminiscent of~\cite{gahinet_apkarian} together with a more general elimination lemma. The solution presented herein is completely different from~\cite{scherer} and may be thought of as a generalization of~\cite{dgkf,lessard_tac} in that explicit formulas for the optimal controller are found. Both approaches are further compared in Section~\ref{sec:LMI_compare}.

In the remainder of this section, a summary of notation and conventions is given. The main result and a short discussion are presented in Section~\ref{sec:main}. Implementation details are given in Section~\ref{sec:example}, and an outline of the proof is given in Section~\ref{sec:proof}.

\paragraph{Common sets and operators.}
Let $\R$ and $\C$ denote the real and complex numbers respectively. $Z^\tp$ and $Z^*$ denote the transpose and conjugate transpose of $Z$, respectively. $\bar\sigma(Z)$ is the maximum singular value and $\rho(A)$ is the spectral radius. The imaginary axis is $j\R$. Let 
$\Ltwo^{m\times n}(j\R)$ be the set of functions $\mathcal{F}:\C\to\C^{m\times n}$ such that the integral
\[
\|\mathcal{F}\|_2^2 \defeq \frac{1}{2\pi}\int_{-\infty}^\infty \trace\bigl(\mathcal{F}(j\omega)^* \mathcal{F}(j\omega)\bigr)\mathrm{d}\omega
\]
is bounded. The subspace $\Htwo^{m\times n}(j\R) \subset \Ltwo^{m\times n}(j\R)$ denotes the functions that are analytic in the open right-half plane. The shorthand $\Htwo$ and $\Ltwo$ is used for brevity. The sets $\Hinf$ and $\Linf$ are similarly defined, but with
\[
\|\mathcal{F}\|_\infty \defeq \esssup_{\omega\in\R} \bar{\sigma} \bigl(\mathcal{F}(j\omega)\bigr)
\]
Let $\Rp$ be the set of proper rational transfer matrices. Every $\mathcal{G} \in \Rp$ has a state-space realization
	\[
	\mathcal{G} = \stsp{A}{B}{C}{D} \defeq D+C(sI-A)^{-1}B
	\]
If this realization is chosen to be stabilizable and detectable,
then~$\mathcal{G}\in\Hinf$ if and only if~$A$ is Hurwitz,
and~$\mathcal{G}\in\Htwo$ if and only if~$A$ is Hurwitz and $D=0$.
For a thorough introduction to these topics, see~\cite{zdg}.

\paragraph{Structured matrices.}
The following notation is used to specify block-lower-triangular matrices.
\[
\Lower(S,m,n) \defeq \set{\bmat{X_{11} & 0 \\ X_{21} & X_{22}}}{X_{ij} \in S^{m_i\times n_j}}
\]
For such block matrices, also define
\[
E_1 \defeq \bmat{I \\ 0} \quad
E^1 \defeq \bmat{I & 0 \\ 0 & 0}\quad
E_2 \defeq \bmat{0 \\ I} \quad
E^2 \defeq \bmat{0 & 0 \\ 0 & I}
\]
where the block dimensions are to be inferred by context. For example, if $A \in \Lower(S,m,n)$, then $E_2^\tp A E_1 = A_{21}$ and $AE^2A^\tp = E_2 A_{22} A_{22}^\tp E_2^\tp$. 

\paragraph{Hamiltonians.}
A Hamiltonian is a matrix of the form
\[
H = \bmat{A & R \\ -Q & -A^\tp}
\]
\if\MODE1\else\newpage\noindent\fi
where $A$ is square and $R$ and $Q$ are symmetric. If $H$ has no eigenvalues on the imaginary axis, and satisfies the complementarity property~\cite{dgkf,zdg}, then $H$ is in the domain of the Riccati operator, written $H\in\domric$. In this case, the associated algebraic Riccati equation (ARE) $A^\tp X + XA + Q + XRX = 0$ has a unique solution such that $A+RX$ is Hurwitz. This stabilizing solution is denoted $X = \ric(H)$, and it is always symmetric.


\section{Main result}\label{sec:main}
Suppose the plant $\mathcal{P}\in\Rp$ has the state-space realization
\begin{equation}
  \label{a:minreal}
  \bmat{ \mathcal{P}_{11} & \mathcal{P}_{12} \\
    \mathcal{P}_{21} & \mathcal{P}_{22} } =
  \left[\begin{array}{c|cc}
      A & B_1 & B_2 \\ \hlinet
      C_1 & 0 & D_{12} \\
      C_2 & D_{21} & 0
    \end{array}\right]
\end{equation}
in which the matrices $A$, $B_2$, $C_2$ have the structure
\begin{equation}\label{a:tri_form}
\begin{gathered}
A \in \Lower(\R,n,n), B_2 \in \Lower(\R,n,m), \\
\text{and } C_2 \in \Lower(\R,k,n).
\end{gathered}
\end{equation}
This ensures that $\mathcal{P}_{22}$ has the requisite block-lower-triangular structure~\eqref{triangular_structure}. The converse is also true; whenever $\mathcal{P}_{22}$ satisfies~\eqref{triangular_structure}, a realization satisfying~\eqref{a:tri_form} can be readily constructed \cite{lessard_realization,voulgaris_stabilization}.
It is further assumed that $A_{11}$ and $A_{22}$ have non-empty dimensions. This avoids trivial special cases and allows for more streamlined results.

Finally, the same assumptions as in~\cite{dgkf} are made on $\mathcal{P}_{22}$ in order to simplify the presentation.
\begin{enumerate}[\bf({A}1)]
\itemsep=1mm
\item $(A,B_1)$ is stabilizable and $(C_1,A)$ is detectable \label{A1}
\item $(A,B_2)$ is stabilizable and $(C_2,A)$ is detectable \label{A2}
\item $D_{12}^\tp \bmat{C_1 & D_{12}} = \bmat{0 & I}$ \label{A3}
\item $D_{21}\bmat{B_1^\tp & D_{21}^\tp} = \bmat{0 & I}$ \label{A4}
\end{enumerate}
As in~\cite{dgkf}, define the Hamiltonians
\begin{equation}\label{def:H_centralized}
\begin{aligned}
H_X &\defeq \bmat{A & \gamma^{-2} B_1B_1^\tp - B_2B_2^\tp \\ -C_1^\tp C_1 & -A^\tp}\\
H_Y &\defeq \bmat{A^\tp & \gamma^{-2} C_1^\tp C_1 - C_2^\tp C_2 \\ -B_1B_1^\tp & -A}
\end{aligned}
\end{equation}
For reference, recall the state-space solution of the classical $\Hinf$ problem stated below as Theorem~\ref{thm:dgkf}. Throughout this paper, we refer to the classical problem as \emph{centralized} or \emph{unstructured}, to distinguish it from \PROB.
%
\begin{thm}[DGKF~\cite{dgkf}]\label{thm:dgkf}
  Suppose $\mathcal{P}\in\Rp$ satisfies~\eqref{a:minreal} as well as Assumptions \bref{A1}--\bref{A4}. There exists a controller that satisfies~\bref{R1}--\bref{R3} if and only if
\begin{enumerate}[\bf({B}1)]
\item $H_X\in\domric$, and $X \defeq \ric(H_X) \geq 0$ \label{B1}
\item $H_Y\in\domric$, and $Y \defeq \ric(H_Y) \geq 0$ \label{B2}
\item $\rho(XY) < \gamma^2$ \label{B3}
\end{enumerate}
  When these conditions hold, one such controller is
  \begin{equation}\label{Kcen}
  \mathcal{K}_\textup{cen} = \stsp{\hat A}{-ZL}{K}{0}
  \end{equation}
where the following definitions were used.
\begin{equation}\label{def:XYZ}
\begin{aligned}
\hat A &\defeq A+B_2 K + ZLC_2 + \gamma^{-2} B_1 B_1^\tp X \\
Z &\defeq (I-\gamma^{-2} YX)^{-1}\\
K &\defeq -B_2^\tp X \quad\text{and}\quad L \defeq -Y C_2^\tp
\end{aligned}
\end{equation}
\end{thm}

The solution to the structured $\Hinf$ problem involves a new pair of Hamiltonians. For any $\hat Y\ge 0$ such that $\rho(X\hat Y) < \gamma^2$, define
\begin{equation}\label{def:J_X}
J_X(\hat Y) \defeq \bmat{A_X & R_X \\
-K^\tp E^1 K & -A_X^\tp }
\end{equation}
where the following definitions were used.
\begin{equation}\label{help1}
\begin{aligned}
A_X &\defeq A + B_2 E^2 K + Z_L \hat L C_2 + \gamma^{-2} B_1 B_1^\tp X \\
R_X &\defeq \gamma^{-2}( B_1 B_1^\tp + Z_L\hat L \hat L^\tp Z_L^\tp ) - B_2 E^2 B_2^\tp \\
\hat L &\defeq -\hat YC_2^\tp E^1 \quad\text{and}\quad Z_L \defeq (I-\gamma^{-2}\hat YX)^{-1}
\end{aligned}
\end{equation}
Note that $Z_L$ is invertible because $\rho(X\hat Y) < \gamma^2$. Similarly, for any $\hat X \geq 0$ such that $\rho(\hat X Y) < \gamma^2$, define
\begin{equation}\label{def:J_Y}
J_Y(\hat X) \defeq \bmat{ A_Y^\tp & R_Y \\
-L E^2 L^\tp & -A_Y}
\end{equation}
where the following definitions were used.
\begin{equation}\label{help2}
\begin{aligned}
A_Y &\defeq A + B_2 \hat K Z_K + L E^1 C_2  + \gamma^{-2} Y C_1^\tp C_1  \\
R_Y &\defeq \gamma^{-2}(C_1^\tp C_1 + Z_K^\tp \hat K^\tp \hat K Z_K) - C_2^\tp E^1 C_2 \\
\hat K &\defeq -E^2 B_2^\tp \hat X  \quad\text{and}\quad 
Z_K \defeq (I-\gamma^{-2}Y\hat X)^{-1}
\end{aligned}
\end{equation}
Note that the Hamiltonians $J_X$ and $J_Y$ are of the standard $\Hinf$ type just like $H_X$ and $H_Y$. That is, the constant term is positive semidefinite, while the quadratic term may be indefinite. The main result is given below. 

\begin{thm}\label{thm:main} 
  Suppose $\mathcal{P}\in\Rp$ satisfies~\eqref{a:minreal}--\eqref{a:tri_form} as well as Assumptions \bref{A1}--\bref{A4}. There exists a controller that solves \PROB~if and only if
  \begin{enumerate}[\bf({C}1)]
  \itemsep=1mm
  \item Conditions \bref{B1}--\bref{B3} hold. In other words, the unstructured version of the problem has a solution.\label{C1}
  \item There exists $\hat X$ and $\hat Y$ such that\\[1mm]
  ${\hspace{-5mm}\begin{cases}
  J_X(\hat Y) \in\domric\text{, and }\hat X - X = \ric( J_X(\hat Y))\geq 0 \\
  J_Y(\hat X) \in\domric\text{, and }\hat Y - Y = \ric( J_Y(\hat X))\geq 0
  \end{cases}}
  $ where both $\rho(X\hat Y) < \gamma^2$ and $\rho(\hat X Y) < \gamma^2$. \label{C2}
  \end{enumerate}
  When these conditions hold, one such controller is
\begin{equation}\label{Kme}
\mathcal{K}_\textup{me} \defeq \left[\begin{array}{cc|c}
\hat A_1 & 0 & -Z_L \hat L \\
B_2(K-\hat KZ_K Z^{-1}) & \hat A_2 & -L \\ \hlinet
K - \hat K Z_K Z^{-1} & \hat K Z_K & 0
\end{array}\right]\vspace{-1mm}
\end{equation}
where\vspace{-2mm}
\begin{align*}
\hat A_1 &\defeq A+B_2K+Z_L\hat L C_2 +\gamma^{-2}B_1B_1^\tp X\\
\hat A_2 &\defeq A+B_2\hat K Z_K + LC_2 +\gamma^{-2} YC_1^\tp C_1
\end{align*}
and $K$, $\hat K$, $L$, $\hat L$, $Z$, $Z_K$, $Z_L$ are defined in~\eqref{def:H_centralized}--\eqref{help2}.
\end{thm}

An outline of the proof is provided in Section~\ref{sec:proof}.
An immediate concern with Theorem~\ref{thm:main} is that there is no obvious way to verify condition \bref{C2}, as it requires solving two intricately coupled AREs. This point is discussed extensively in Section~\ref{sec:example}, where an efficient numerical method is proposed that can be used to find a fixed point when it exists. In the remainder of this section, some salient features of the optimal controller are discussed.


\paragraph{Coordinates.} Let $\xi$ be the state used in the realization of the optimal unstructured controller \eqref{Kcen} in Theorem~\ref{thm:dgkf}. If the state $\zeta \defeq Z^{-1} \xi$ is used instead, the following dual realization is obtained.
\begin{equation}\label{Kcen2}
\!\!\mathcal{K}_\textup{cen} = \stsp{A+B_2 KZ + LC_2 + \gamma^{-2} Y C_1^\tp C_1}{-L}{KZ}{0}
\end{equation}
Likewise, there are many possible coordinate choices for representing the optimal structured controller~\eqref{Kme}. 

In the closely related treatment of risk-sensitive optimal control by Whittle~\cite{whittle}, the state $\xi$ has the interpretation of extremizing \emph{total stress}, while $\zeta$ extremizes \emph{past stress}. In Theorem~\ref{thm:main}, the controller is expressed in \emph{mixed} coordinates $(\xi^1,\zeta^2)$. That is, a $\xi$-like coordinate for the first state and a $\zeta$-like coordinate for the second.
This choice was made because it yields the simplest-looking formulae for $\mathcal{K}_\textup{me}$. Coordinate choice is further discussed in the proof outline in Section~\ref{sec:proof}.
\paragraph{Structure.} The controller $\mathcal{K}_\textup{me}$ may also be written in the standard observer form. The first state equation is
\begin{align}\label{eqq1}
\dot{\xi}^1 = A \xi^1 + B_1 \hat w^1 + B_2 \hat u^1 - Z_L\hat L (y - C_2 \xi^1)
\end{align}
where $\hat w^1 \defeq \gamma^{-2}B_1^\tp X$ and $\hat u^1 \defeq K \xi^1$ are precisely the \emph{worst-case noise} and \emph{optimal input} respectively in the classical case~\cite{dgkf}. In fact,~\eqref{eqq1} is identical to the classical centralized estimator except $Y$ has been replaced by $\hat Y$.
The second state equation is
\begin{align}\label{eqq2}
\dot{\zeta}^2 = A \zeta^2 + B_2 u +\gamma^{-2}YC_1^\tp C_1 - L (y - C_2 \zeta^2)
\end{align}
which is the estimator from~$\mathcal{K}_\textup{cen}$ expressed in the $\zeta$-coordinates, but with the optimal two-player $u$ from Theorem~\ref{thm:main} rather than the centralized $u = K\xi$ from Theorem~\ref{thm:dgkf}. These structural properties suggest that~$\mathcal{K}_\textup{me}$ exhibits a separation structure similar to the one described in~\cite{dgkf}, but further work is needed to state it precisely.

\paragraph{Limiting behavior.} As mentioned in Section~\ref{sec:intro}, entropy tends to the squared $\Htwo$-norm in the limit $\gamma\to\infty$. When this limit is considered for the DGKF controller, then $Z= I$, $\zeta = \xi$ and the two realizations of $\mathcal{K}_\textup{cen}$ from~\eqref{Kcen} and~\eqref{Kcen2} coincide.

Now consider the limit $\gamma\to\infty$ for $\mathcal{K}_\textup{me}$ in~\eqref{Kme}. Then $Z = Z_L = Z_K = I$ and the $\Htwo$-optimal structured controller from~\cite{lessard_lall_tpof} is recovered:
\[
\mathcal{K}_\textup{rn} \defeq \left[\begin{array}{cc|c}
A+B_2K+\hat L C_2 & 0 & -\hat L \\
B_2(K-\hat K) & A+B_2\hat K + LC_2 & -L \\ \hlinet
K - \hat K  & \hat K  & 0
\end{array}\right]
\]
It can also be shown that in the limit $\gamma\to\infty$, the complicated condition~\bref{C2} simplifies to the simple linearly coupled equations given in~\cite{lessard_lall_tpof}.


\section{Comparison with LMI method}\label{sec:LMI_compare}

The only existing solution to \PROB~is the work by Scherer~\cite{scherer}, which finds an LMI characterization of the $\gamma$-suboptimal controllers. Lower-triangular structures with more than two players are also considered in~\cite{scherer}. As in the classical case, there are many benefits to using an LMI approach; for example it allows a seamless treatment of singular problems~\cite{gahinet_apkarian}.
The DGKF solution~\cite{dgkf} makes more assumptions and is therefore more limited in its applicability. However, the DGKF solution provides observer-based formulas that give a clear and powerful interpretation of the controller's role and structure.

With regards to computational complexity, the DGKF solution is more efficient than the LMI approach. If $A\in \R^{n\times n}$, then the complexity of testing {\bf(B1)--(B3)} is dominated by solving two AREs and finding one spectral radius. These are essentially eigenvalue problems and can be solved in $\ord(n^3)$. The LMI formulation results in a semidefinite program (SDP) with two $n\times n$ decision variables. It therefore has a complexity of $\ord(n^6)$ when using a conventional interior-point method.

For \PROB, the LMI test by Scherer~\cite{scherer} has more variables than the centralized LMI solution, but nevertheless has complexity $\ord(n^6)$.  Verifying the conditions in Theorem~\ref{thm:main} is dominated by the task of finding $\hat X$ and $\hat Y$ that satisfy~\bref{C2}. To this end, a simple and efficient algorithm is given in Section~\ref{sec:example} that roughly amounts to iteratively solving each ARE until convergence is achieved. Each step has complexity $\ord(n^3)$ and it is verified empirically that convergence to machine precision takes fewer than 15 iterations and is independent of $n$.

Despite the proposed iterative ARE method being more computationally efficient than the LMI approach, both involve a necessary and sufficient condition for~\PROB. One therefore expects there to exist a transformation of the coupled AREs of Theorem~\ref{thm:main} into the LMI condition of~\cite{scherer} and vice versa. Such a construction for the centralized case is detailed in~\cite{gahinet_followup}, but is the subject of future work for the unstructured case.


\section{Iterative solution}\label{sec:example}

As mentioned in Section~\ref{sec:main}, it is not clear how one would verify \bref{C2} in Theorem~\ref{thm:main}. In this section, preliminary results are presented that suggest that a simple iterative scheme may be used to efficiently verify \bref{C2}.

\paragraph{Iterative scheme (\ITS).}\label{ITERATE} Given some $\gamma > 0$ and a starting guess $\hat Y_0$, solve the following AREs iteratively
\begin{equation}\label{iter}
\begin{aligned}
\hat X_{k+1} &= X + \ric( J_X(\hat Y_k) ) \\
\hat Y_{k+1} &= Y + \ric( J_Y(\hat X_{k+1}) )
\end{aligned}
\quad\text{for }k=0,1,\dots
\end{equation}
and stop when $\hat X_k$ and $\hat Y_k$ have converged to some $\hat X$ and~$\hat Y$ respectively. Then check to see if $\hat X\geq X$, $\hat Y\geq Y$, $\rho(X\hat Y) < \gamma^2$, and $\rho(\hat X Y) < \gamma^2$. If so, then \bref{C2} is verified. If these conditions are not met, or if $J_X \notin \domric$ or $J_Y \notin\domric$ for any of the iterates, then the test is inconclusive.

\paragraph{Rapid convergence.} The main issue with~\ITS~is choosing a suitable initial point. In other words, finding $\hat Y_0$ such that $J_X(\hat Y_0)\in\domric$. This task becomes increasingly difficult as $\gamma$ approaches $\gamma_\textup{opt}$, the infimum over all $\gamma$ that solves \PROB. When $\gamma\to\infty$, \bref{C2} is satisfied by the $\Htwo$ values of $\hat X$ and $\hat Y$. Therefore, these limiting values of $\hat X$ and $\hat Y$, which are easily computed as in~\cite{lessard_lall_tpof,lessard_tac}, are good initializations for~\ITS~when $\gamma$ is sufficiently large.

\begin{figure}[ht]
	\centering
	\begin{tikzpicture}
	\begin{semilogyaxis}[
		enlarge x limits={abs value=0,auto},
		cycle list name=gray lines,
		xlabel={Iteration number $k$},
		xmax = 19,
		ymax = 0.1,
		ytick={1e-2,1e-4,1e-6,1e-8,1e-10,1e-12,1e-14,1e-16}
	]
	\pgfplotstableread{N20.txt}\table
  		\foreach \y in {1,...,100}
 			\addplot table[x index=0, y index=\y] from \table;
	\end{semilogyaxis}
	\end{tikzpicture}
\caption{Convergence error $e_k$ for the \ITS~algorithm. 100 random systems with $n=20$ states at $\gamma\approx 2\gamma_\text{opt}$.\label{fig:plot_err}}
\end{figure}
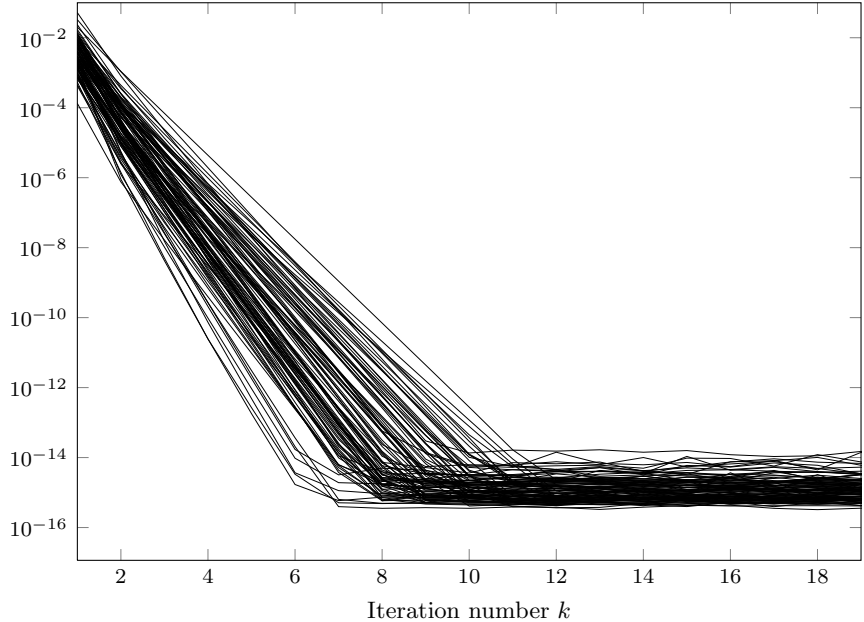

To investigate this initialization, random structured systems with $n$ states, and  $\frac{n}{2}$ inputs and outputs were generated. The MATLAB function \texttt{rss} was used to generate $(A_{11},B_{11},C_{11})$ and $(A_{22},B_{22},C_{22})$, while \texttt{randn} was used to generate $A_{21}$, $B_{21}$, $C_{21}$, $B_1$, $C_1$. Matrices $D_{12}$ and $D_{21}$ were chosen to satisfy \bref{A3}--\bref{A4}. Finally, $B_1$ and $C_1$ were each scaled by $1/\sqrt{n}$. The result was a family of systems for which the infimal  centralized $\gamma_\textup{cen}$ is approximately $3$ for all $n$. For each test, $\gamma_\text{opt}$ was approximated using the LMI method~\cite{scherer} and then \ITS~was performed for $\gamma = 2\gamma_\text{opt}$ using the initialization described above. 100 tests were performed for each $n\in\{4,8,12,16,20\}$. Valid $\hat X$ and $\hat Y$ satisfying \bref{C2} were successfully obtained in every case. In Figure~\ref{fig:plot_err}, the convergence error
\[
e_k \defeq \frac1n\sqrt{\|\hat X_k-\hat X\|_F^2 + \|\hat Y_k- \hat Y\|_F^2}
\]
is plotted a function of the iteration $k$ for the case $n=20$.

No appreciable difference in convergence rate was observed as the state dimension $n$ was increased; convergence was achieved in fewer than 15 iterations every time.

Note that \texttt{rss} only produces stable systems. If unstable modes are included in $A$, then even the optimal \emph{centralized} $\gamma_\textup{cen}$ becomes very volatile and can sometimes exceed $10^4$ depending on the number of unstable modes. The convergence of \ITS~is still linear in the presence of instability, but the rate is typically worse and more variable than the stable case, often taking 30--50 iterations to reach convergence. The slowest cases tested took up to 200 iterations. Nevertheless, the performance of~\ITS~still appears to be independent of $n$ as in the stable case.

\paragraph{Warm-start technique.} When $\gamma$ is too close to $\gamma_\textup{opt}$, initializing $\hat Y_0$ with the limiting value of $\hat Y$ is sometimes ineffective. One possible solution is to iteratively decrease $\gamma$ and set $\hat Y_0$ to be the converged $\hat Y$ from the previous $\gamma$ iteration. Preliminary simulations indicate that this warm-started approach works very well, and $\gamma_\text{opt}$ is achieved as long as $\gamma$ is not decreased too rapidly.


\section{Outline of the proof}\label{sec:proof}

The proof of Theorem~\ref{thm:main} is algebraically involved, but conceptually simple. Due to space constraints, an outline of the proof is given that highlights the key enabling insights. The conditions \bref{C1}--\bref{C2} are necessary and sufficient for there to exist a solution to \PROB; each direction is addressed separately.

\paragraph{Proof of sufficiency.} Suppose that \bref{C1}--\bref{C2} hold. The aim is to verify \bref{R1}--\bref{R4} separately.

It is immediate that \bref{R4} is satisfied because each of the state-space matrices in~\eqref{Kme} is block-lower-triangular. This follows from the sparsity of $\hat L$ and $\hat K$.

Requirements \bref{R1}--\bref{R2} are verified by appealing to the bounded real lemma and a well-known relationship between an ARE its associated algebraic Riccati inequality (ARI). We state these results as lemmas.
\begin{lem}[bounded real lemma] \label{lem:brl}
Suppose $\mathcal{G}$ has realization $(A,B,C,0)$ and $\gamma > 0$ is given. The following statements are equivalent.
\begin{enumerate}[(i)]
\item $A$ is Hurwitz and $\|\mathcal{G}\|_\infty < \gamma$
\item There exists $X > 0$ such that\\
$
A^\tp X + X A  + \gamma^{-2} X B B^\tp X + C^\tp C< 0
$
\end{enumerate}
\end{lem}

\begin{lem}\label{lem:are_ari}
Suppose $R$ and $Q$ are symmetric and either $R\geq 0$ or $Q\geq 0$. Then the following are equivalent
\begin{enumerate}[(i)]
\item There exists $X > 0$ satisfying the inequality\\$A^\tp X + XA + X R X+ Q  < 0$.
\item There exists $X_0 \geq 0$ such that $A+RX_0$ is Hurwitz and
$A^\tp X_0 + X_0A + X_0RX_0 + Q = 0$
\end{enumerate}
If the above conditions hold, then $0 \leq X_0 < X$.
\end{lem}

Applying Lemma~\ref{lem:brl} to the closed-loop map $\cl$~\eqref{closed-loop_map} induced by the proposed~$\mathcal{K}_\textup{me}$, we find {\it (i)} is equivalent to \bref{R1}--\bref{R2}. Rather than solving the inequality in Lemma~\ref{lem:brl}, a stabilizing solution to the associated ARE is constructed and Lemma~\ref{lem:are_ari} is applied.

There are many possible realizations for $\cl$ to choose from. If the plant $\mathcal{P}$ has state $x$, and $\mathcal{K}_\textup{me}$ has states $(\xi^1,\zeta^2)$, then define the coordinate choices
\begin{align*}
m&:\bmat{\xi^1 \\ x-\xi^1 \\ x-\zeta^2}  &
x&:\bmat{x\\x-\xi^1 \\ x-\xi^2} &
y&:\bmat{\zeta^1\\\zeta^2-\zeta^1\\x-\zeta^2}
\end{align*}
The new coordinates $\xi^2$ and $\zeta^1$ are related to the states of $\mathcal{K}_\textup{me}$ as follows.
\[
\bmat{\zeta^1 \\ \zeta^2}
= \bmat{ Z_L^{-1} & 0 \\ Z^{-1} - Z_K^{-1} & Z_K^{-1} } \bmat{\xi^1 \\ \xi^2}
\]
This relationship is a generalization of the classical~$\Hinf$ coordinate transform $\zeta = Z^{-1} \xi$ mentioned in Section~\ref{sec:main}. The $x$ and $y$ coordinates were chosen because they yield simple solutions to the bounded real equation. The result is given by the following lemma.
\begin{lem}\label{lem:big_are_verify}
Consider the setting of Theorem~\ref{thm:main}. Express the closed-loop map $\cl$ in the $x$ and $y$ coordinates, and associate the following labels to the corresponding state-space matrices.
\begin{equation}\label{eq:CLcoords}
\cl = \stsp{\bar A_x}{\bar B_x}{\bar C_x}{0} = \stsp{\bar A_y}{\bar B_y}{\bar C_y}{0}
\end{equation}
Define the associated Hamiltonians
\begin{equation*}
\bar H_X \defeq \shrink{5.5}\bmat{ \bar A_x & \gamma^{-2}\bar B_x \bar B_x^\tp \\[1mm]
-\bar C_x^\tp \bar C_x & -\bar A_x^\tp}\!\!,\,
\bar H_Y \defeq \bmat{ \bar A_y^\tp & \gamma^{-2}\bar C_y^\tp \bar C_y \\[1mm]
-\bar B_y \bar B_y^\tp & -\bar A_y}
\end{equation*}
Then $\bar H_X \in \domric$ and $\bar H_Y \in \domric$. Moreover,
\[
\ric(\bar H_X) =\! \shrink{2}\bmat{X & 0 & 0 \\ 0 & \hat X \!-\! X & 0 \\ 0 & 0 & \Phi}\!\!,\,
\ric(\bar H_Y) =\! \bmat{\Psi & 0 & 0 \\ 0 & \hat Y \!-\! Y & 0 \\ 0 & 0 & Y}
\]
where $X$, $Y$, $\hat X$, $\hat Y$ are defined in \bref{B1}--\bref{B2} and \bref{C2}, and $\Phi\geq 0$ and $\Psi\geq 0$.
\end{lem}
The variables $\Phi$ and $\Psi$ from Lemma~\ref{lem:big_are_verify} are also solutions to AREs, but the associated formulas are omitted because the values of $\Phi$ and $\Psi$ are unimportant.

By Lemma~\ref{lem:are_ari}, the solutions to the AREs given in Lemma~\ref{lem:big_are_verify} imply that the bounded real inequalities also have solutions, and so by Lemma~\ref{lem:brl}, $\cl$ has a stable state-space realization and $\|\cl\|_\infty < \gamma$. In other words, requirements \bref{R1}--\bref{R2} have been verified.

Verifying \bref{R3}, or that the proposed controller minimizes entropy, is accomplished by first deriving a necessary and sufficient condition for optimality and then checking that it is satisfied by the proposed $\mathcal{K}_\textup{me}$.
\newpage
\begin{lem}
\label{lem:opt_condition}
Suppose the set of admissible closed-loop maps is parameterized by
\[
\cl \in \set{\mathcal{T}_1 + \mathcal{T}_2 \mathcal{Q} \mathcal{T}_3}{\mathcal{Q}\in\Lower(\Htwo,m,k)}
\]
where the $\mathcal{T}_i$ are stable.
If $\|\cl\|_\infty < \gamma$, then 
$\cl$ has minimum entropy if and only if
\begin{equation}\label{optcond}
\mathcal{T}_2^* \cl (I - \gamma^{-2}\cl^*\cl)^{-1}\mathcal{T}_3^* \in \bmat{\Htwo^\perp & \Ltwo \\ \Htwo^\perp & \Htwo^\perp}
\end{equation}
\end{lem}
A similar approach was taken in~\cite{lessard_tac} to prove optimality in the $\Htwo$ case. Indeed, in the limit $\gamma\to\infty$, Equation \eqref{optcond} becomes the $\Htwo$ optimality condition from~\cite{lessard_tac}.

It has already been verified that $\|\cl\|_\infty < \gamma$, so it remains to show that the optimality condition holds. The closed-loop map may be expressed in the affine form $\cl = \mathcal{T}_1 + \mathcal{T}_2 \mathcal{Q}\mathcal{T}_3$ using a modified Youla parameterization. The following result is from~\cite{lessard_tac}. Similar parameterizations were also reported in~\cite{lessard_realization} and~\cite{voulgaris_stabilization}.

\begin{lem}\label{lem:param}
Suppose $\mathcal{P}$ and its state-space realization satisfies~\eqref{a:minreal}--\eqref{a:tri_form}. There exists $\mathcal{K}\in\Lower(\Rp,m,k)$ that stabilizes $\mathcal{P}$ if and only if $(C_{ii},A_{ii},B_{ii})$ is stabilizable and detectable for $i=1,2$.
In this case, let $K_i$ and $L_i$ be such that $A+B_{ii}K_i$ and $A+L_iC_{ii}$ are Hurwitz. Then define $K_d \defeq \diag(K_1,K_2)$ and $L_d\defeq \diag(L_1,L_2)$. the set of all stabilized closed-loop maps is parameterized by
\begin{equation}\label{eq:Tclparam}
\cl \in \set{\mathcal{T}_1 + \mathcal{T}_2 \mathcal{Q} \mathcal{T}_3}{ \mathcal{Q}\in\Lower(\Htwo,m,k)}
\end{equation}
where the $\mathcal{T}_i$ matrices have the joint realization
\begin{equation}
    \label{eqn:T}
    \bmat{\mathcal{T}_{1} & \mathcal{T}_{2} \\ \mathcal{T}_{3} & 0} =
    \left[\begin{array}{cc|cc}
        A_{Kd} &  -B_2 K_d & B_1 & B_2 \\
        0 & A_{Ld} & B_{Ld} & 0 \\ \hlinet
        C_{Kd} & -D_{12}K_d & 0 & D_{12} \\
        0 & C_2 & D_{21} & 0
      \end{array}\right]
  \end{equation}
and the following shorthand notation was used.
\begin{equation}\label{defn:short}
\begin{aligned}
A_{Kd} &\defeq A+B_2K_d 		& A_{Ld} &\defeq A+L_dC_2 \\
C_{Kd} &\defeq C_1+D_{12}K_d 	& B_{Ld} &\defeq B_1+L_dD_{21}
\end{aligned}
\end{equation}
\end{lem}
There is also a one-to-one mapping between each stabilizing controller $\mathcal{K}$ and its associated $\mathcal{Q}$-parameter, but these details are left out of Lemma~\ref{lem:param} to save some space.
Using the parameterization of Lemma~\ref{lem:param}, the optimality condition~\eqref{optcond} may be verified by direct substitution and appropriate state-space simplifications.

\paragraph{Proof of necessity.} This part of the proof is similar to how necessity was proved in the $\Htwo$ case~\cite{lessard_tac}. Roughly, the $\mathcal{Q}_{11}$ part of the controller from the parameterization of Lemma~\ref{lem:param} is held fixed and the problem of finding the minimum-entropy $\bmat{\mathcal{Q}_{21} & \mathcal{Q}_{22}}$ is considered. This problem is unstructured so  Theorem~\ref{thm:dgkf} may be applied. The result is that a pair of AREs must have positive-semidefinite solutions and a spectral radius condition must be met. After some algebraic manipulations, it is found that the AREs are those that correspond to the Hamiltonians $J_X$ and $H_Y$ and the spectral radius condition amounts to $\rho(\hat X Y) < \gamma^2$. Using a similar argument, holding $\mathcal{Q}_{22}$ fixed leads to the conditions on $J_Y$ and $H_X$ together with $\rho(X \hat Y) < \gamma^2$.


\section{Acknowledgments}
The author would like to thank Bo Bernhardsson, \mbox{Sanjay Lall}, Andrew Packard, and Peter Seiler for helpful discussions. This work was supported in part by NASA under Grant No. NRA NNX12AM55A.
\vspace{-1mm}


\if\MODE1\else\begin{small}\fi
\bibliographystyle{abbrv}
\bibliography{hinfinity}
\if\MODE1\else\end{small}\fi

\end{document}